\documentclass[sigconf,nonacm]{aamas}

\usepackage{balance} 

\setcopyright{ifaamas}
\acmConference[AAMAS '26]{Proc.\@ of the 25th International Conference
on Autonomous Agents and Multiagent Systems (AAMAS 2026)}{May 25 -- 29, 2026}
{Paphos, Cyprus}{C.~Amato, L.~Dennis, V.~Mascardi, J.~Thangarajah (eds.)}
\copyrightyear{2026}
\acmYear{2026}
\acmDOI{}
\acmPrice{}
\acmISBN{}

\usepackage{tabularx}
\usepackage{subcaption}

\acmSubmissionID{1256}

\theoremstyle{definition}

\usepackage{booktabs,threeparttable,siunitx}
\newcommand{\sym}[1]{\ifmmode^{#1}\else\(^{#1}\)\fi}

\title{Responsibility in Multi-Agent Sequential Decision-Making: Comparing Human Judgments to Formal Models of Causal Attribution}

\author{Nripsuta Ani Saxena}
\authornote{Work performed during an internship at the Max Planck Institute for Software Systems.}
\affiliation{%
  \institution{University of Southern California}
  \city{Los Angeles}
  \state{California}
  \country{USA}
}
\email{nsaxena@usc.edu}

\author{Stelios Triantafyllou}
\affiliation{%
  \institution{Max Planck Institute for Software Systems}
  \city{Saarbrücken}
  \country{Germany}
}
\email{strianta@mpi-sws.org}

\author{Goran Radanović}
\affiliation{%
  \institution{Max Planck Institute for Software Systems}
  \city{Saarbrücken}
  \country{Germany}
}
\email{gradanovic@mpi-sws.org}

\begin{document}

\begin{abstract}
With the growing adoption of artificial intelligence in high-stakes decision-making domains, identifying the causes of outcomes--particularly failures--and determining who is responsible has become a critical concern. In this work, we investigate how well formal definitions of \textit{responsibility attribution}, grounded in the framework of \textit{actual causality}, align with human judgments of responsibility. 
To this end, we conduct a large-scale survey to elicit human judgments of responsibility in multi-agent sequential decision-making scenarios, using a modified version of the card game Goofspiel. We evaluate different responsibility attribution methods, assessing their alignment with human judgments about responsibility, and identifying factors that significantly shape responsibility judgments. While no single responsibility attribution method consistently aligns with human responses, our findings highlight key factors that influence human responsibility judgments, including agent-specific biases and amount of information available to agents during decision-making.
\end{abstract}

\maketitle


\section{Introduction}\label{sec.introduction}

In high-stakes multi-agent sequential decision-making, attributing responsibility when harm occurs is a central concern for ensuring accountable outcomes. This task is inherently complex due to several intertwined challenges. Uncertainty about the consequences of individual and collective decisions makes it difficult to assess causal links between actions and outcomes. Furthermore, multi-agent settings face the {\em problem of many hands}~\cite{van2015moral}, where the actions of multiple agents interact in ways that blur the lines of individual responsibility. Temporal interdependence between decisions---where decisions made by one agent influence and are influenced by the actions of others over time---further complicates attribution. These factors imply that assigning responsibility cannot rely on simple causal or temporal proximity but instead requires careful modeling of agent interactions and their evolving contributions to outcomes.

Recent work has attempted to tackle these challenges by providing a formal framework, grounded in actual causality, for attributing responsibility in complex multi-agent sequential decision-making environments~\cite{triantafyllou2022actual, triantafyllou2023towards}. This framework treats the degree of responsibility as a quantitative measure of the extent to which an agent contributed to an outcome. The central idea is to identify the actual causes of the outcome, i.e., the specific decisions that were pivotal in bringing it about, and then determine an agent's responsibility based on whether their decisions appear among these causes and how significantly they contributed. This approach is axiomatic and prescriptive: it begins with formal principles that define actual causation and systematically derives responsibility assignments by translating causal influence into measurable quantities.

However, these causal approaches to responsibility attribution do not incorporate human factors, making it unclear how well they align with human judgments about responsibility. Yet, as argued by Lima et al.~\cite{lima2021conflict,lima2023blaming}, public opinion matters, especially when decisions have substantial societal impact. Hence, in this paper, we ask: 
\begin{enumerate}
    \item {\em How do 
    responsibility attribution methods based on actual causality align with human judgments about responsibility?}  
\item {\em Which factors most influence how humans assign responsibility in complex multi-agent scenarios?}
\end{enumerate}
To this end, we design a human-subjects study and conduct a survey with 640 respondents to elicit human judgments about responsibility in multi-agent sequential decision-making. We use a team-based version of the card game Goofspiel~\cite{lanctot2019openspiel,triantafyllou2022actual} as a representative setting for complex multi-agent scenarios. The game involves two teams, Team A and Team B, each composed of one human and one AI player (agent), making decisions over five rounds. In our survey, respondents are shown various game scenarios---{\em vignette}---in which Team A {\em loses} and are asked to attribute responsibility for the loss among the members of Team A. In other words, respondents are asked to assign a degree of responsibility to each player in Team A for the team losing in each vignette---shown to them.
In addition, the respondents report their confidence in their assessment.

Our survey design includes multiple combinations of vignette conditions, enabling us to test the alignment of human judgments with various responsibility attribution methods found in the literature. It also allows us to quantify the influence of different factors that might affect human judgment. Specifically, we considered five different responsibility attribution methods based on three definitions of actual causality and two definitions of the degree of responsibility. We focus on three main factors that might influence human judgments, listed below:

 \begin{itemize}
    \item {\bf Initial condition}: Initial conditions were controlled through the initial cards dealt to the players. In some matches, Team A, or a member of Team A, was dealt worse cards than the other players, i.e., there was a {\em bias} against them. Our hypothesis was that people would assign a lower degree of responsibility to a player when they were given a disadvantageous initial condition. 
    Furthermore, by using vignettes where only one of the players of Team A was subject to bias, we test whether respondents distinguish between human and AI agents under unequal starting conditions. 
    
    \item {\bf State observability}: 
    We asked respondents to assign responsibility assuming the games were either {\em perfect-information} (i.e., players see each other’s cards) or {\em imperfect-information} (i.e., players see only their own cards).
    Our hypothesis was that coordination under partial observability is more difficult, and therefore, people would assign lower degrees of responsibility in those scenarios.
    \item {\bf Aid in counterfactual reasoning}: Since the responsibility attribution methods of interest are based on actual causality, counterfactual reasoning plays a critical role. We tested how human judgments differ when counterfactuals are provided, that is, when respondents are shown what would have happened if players from Team A had made different decisions. Our hypothesis was that counterfactuals help align human judgments more closely with the attribution methods that rely on those same counterfactuals.
 \end{itemize}

 \subsection{Overview of Results}
 Our findings show that human judgments of agent responsibility are context-sensitive. Greater information access and the presence of counterfactuals both raise perceived accountability. That is, respondents assigned higher degrees of responsibility to players when they fully observed the state of the game and when counterfactuals were provided to the respondents. This suggests that people calibrate their judgments based on the information available to the agents and what the agents could plausibly do. Notably, under biased initial conditions, respondents assigned higher degrees of responsibility when the disadvantaged player was a human, but not when it was an AI. This asymmetric effect suggests that people calibrate responsibility not only on the basis of an agent’s actions, but also on the agent’s identity.

When it comes to the alignment of respondents’ responses with formal models of causal attribution, it was strongest when the initial conditions were biased, particularly when they were biased against both players in Team A. This suggests that visible structural asymmetries (e.g., biased hands in our experiments) prompt people to assess responsibility in ways that are more consistent with formal models. In contrast, variations in state observability, the number of counterfactuals presented, and the definitions of actual causality and the degree of responsibility did not significantly influence the alignment, indicating no systematic preference for any particular responsibility attribution method.

Finally, we observed that respondents' self-reported confidence was positively correlated with responsibility ratings and alignment to formal models of causal attribution. Respondents who reported higher confidence were more likely to assign higher degrees of responsibility and showed better alignment with the formal responsibility attribution methods. This suggests that confidence reflects a sense of causal clarity.

Our work contributes to the growing body of literature 
on understanding human judgments about responsibility in multi-agent settings involving humans. In contrast to prior work, we focus on examining the alignment between human judgments of responsibility and formal approaches to responsibility attribution, grounded in causality. Furthermore, we consider more complex environments than prior work (e.g., \cite{lima2021human}); in our setting, decision-makers make a sequence of decisions over multiple rounds.


\section{Background}\label{sec.background}

We consider responsibility attribution (RA) methods that first identify actual causes of an outcome using one of the definitions of actual causality from the AI literature~\cite{halpern2016actual, triantafyllou2022actual}. In particular, we consider three definitions of actual causation (\textsc{AC}): the But-For definition (henceforth, the \textsc{BF} definition)~\cite{hart1985causation}, the Halpern and Pearl definition (henceforth, the \textsc{HP} definition)~\cite{halpern2005causes, halpern2016actual}, and a definition introduced in~\cite{triantafyllou2022actual} (henceforth, the \textsc{TR} definition).
Once the actual causes of an outcome have been identified, the question becomes: how much responsibility should each agent bear for that outcome? To answer this question, we can define the degree of an agent’s responsibility. We consider two definitions of the degree of responsibility (\textsc{DR}): one due to~\cite{chockler2004responsibility} (henceforth, the \textsc{CH} degree of responsibility) and another due to~\cite{triantafyllou2022actual} (henceforth, the \textsc{TR} degree of responsibility).
Note that the \textsc{CH} and \textsc{TR} degrees of responsibility are agnostic to the specific definition of actual causality---any definition can be used to determine actual causes. In total, this gives six possible \textsc{AC} $\times$ \textsc{DR} combinations. However, two combinations (\textsc{BF} $\times$ \textsc{CH} and \textsc{BF} $\times$ \textsc{TR}) yield the same degrees of responsibility. Hence, in total, we have five distinct responsibility attribution methods.

The following paragraphs explain the core intuitions behind the aforementioned definitions. We begin by describing three definitions of actual causality. To introduce these definitions, we consider a standard multi-agent sequential decision-making setting under uncertainty, in which a group of agents make decisions within an environment that evolves over time. At each time step, the agents simultaneously select actions based on their individual information states. Subsequently, the environment state and the agents’ information states stochastically transition to their respective next states. For a more formal treatment of these definitions within the Dec-POMDP (decentralized partially observable Markov decision process) framework, we refer the reader to~\cite{triantafyllou2022actual}.

\textbf{The \textsc{BF} Definition of Actual Causality.}
The BF definition of actual causality is based on the {\em but-for} test~\cite{hart1985causation}. Under this criterion, a set of agents’ actions is considered a cause of an outcome if the outcome would not have materialized had this set of actions been altered. Importantly, this set should be minimal in the sense that no strict subset of it passes the but-for test. The minimality condition captures the idea of necessity: an action qualifies as part of an actual cause only if it is required to pass the but-for test. Prior work has argued that the BF definition does not suffice for determining causality, providing examples in which the BF definition either fails to identify intuitively expected actual causes~\cite{halpern2016actual} or includes actions that were not actually taken by agents~\cite{triantafyllou2022actual}. To address these issues, more nuanced definitions of actual causality have been proposed in the literature, one of which is due to Halpern and Pearl~\cite{halpern2005causes}.

\textbf{The \textsc{HP} Definition of Actual Causality.}
Since the Halpern and Pearl definition of actual causality has been refined over time~\cite{halpern2005causes,halpern2016actual}, we adopt in our study the most recent variant proposed in~\cite{halpern2015modification}. The \textsc{HP} definition extends the classical \textsc{BF} definition by introducing contingencies. When performing the but-for test for a target set of agents’ actions, one must account for the causal influence that actions in the target set have on other actions outside it. After all, the but-for test relies on counterfactual reasoning: we assess how the world would have looked had the target set been different. This can be formalized through interventions on the action variables in the target set and includes potential changes in actions not in the target set, meaning that their counterfactual values may differ from their factual ones. The \textsc{HP} definition introduces a contingency set to express counterfactual statements in the but-for test where changing the target set does not influence actions in the contingency set, i.e., the actions in the contingency set are fixed to their factual values. Halpern~\cite{halpern2015modification} argues that the \textsc{HP} definition overcomes some of the drawbacks of the \textsc{BF} definition observed in the literature. That said, Triantafyllou et al.~\cite{triantafyllou2022actual} argue that the \textsc{HP} definition can still lead to counterintuitive actual causes, motivating a new definition of actual causation, which is explained in the next subsection.

\textbf{The \textsc{TR} Definition of Actual Causality.}  
The TR definition, introduced in \cite{triantafyllou2022actual}, extends the HP definition in two ways. First, it modifies the minimality condition to include both actual causes and contingencies. Second, it introduces information states to determine what can be part of an actual cause and a contingency in the but-for test. Using examples, Triantafyllou et al. \cite{triantafyllou2022actual} argue that this definition is more suitable for multi-agent sequential decision-making and overcomes some of the challenges faced by the BF and HP definitions in such settings. 

\textbf{The CH Degree of Responsibility.} 
The \textsc{CH} degree of responsibility \cite{chockler2004responsibility} measures how responsible an agent is for an outcome by examining whether any of the agent’s actions are part of the actual causes identified. 
To simplify the exposition, we explain the definitions of the degree of an agent's responsibility without considering contingencies; see \cite{triantafyllou2022actual} for the complete definitions that include them. If none of the agent’s actions are part of any of the actual causes identified, then their \textsc{CH} degree of responsibility is zero. Otherwise, the agent’s degree of responsibility is determined by the actual cause that has the highest percentage of the agent’s actions, and it is equal to this percentage.

\textbf{The TR Degree of Responsibility.} 
The \textsc{TR} degree of responsibility \cite{triantafyllou2022actual} is a modification of the \textsc{CH} degree of responsibility that prevents agents from reducing their own degree of responsibility by increasing the total number of actions that must be changed in order to improve the outcome. See \cite{triantafyllou2022actual} for the complete definition.

\section{Related Work}\label{sec.literature}
We identify three closely related topics:{\em actual causality}, {\em formal models of responsibility}, and {\em moral responsibility} and {\em human perceptions}.

{\bf Actual causality.}
This paper relates to an extensive body of literature on actual causality. The previous sections cover three definitions of actual causality: the BF definition~\cite{hart1985causation}, the HP definition~\cite{halpern2015modification}, and the TR definition~\cite{triantafyllou2022actual}. Much of the prior work on actual causality has focused on extending the BF definition~\cite{pearl1998definition,hitchcock2001intransitivity,hall2007structural,halpern2015graded}. We refer the reader to the related work section of \cite{triantafyllou2022actual}, which provides a brief overview of these definitions and explains how they relate to the HP definition and its variants~\cite{halpern2005causes, halpern2015modification, halpern2016actual}.
Our goal is to study human perceptions in complex multi-agent sequential decision-making. Hence, we focus on the definitions that have been operationalized in one such setting (i.e., in Dec-POMDPs)~\cite{triantafyllou2022actual}. In contrast to prior work that studies the computational and structural properties of these definitions~\cite{triantafyllou2022actual,triantafyllou2023towards}, we aim to understand how responsibility-attribution methods based on these definitions align with human responsibility judgments and the factors that contribute to this alignment.

{\bf Formal models of responsibility.} The most relevant prior work to ours concerns models of responsibility based on actual causality~\cite{chockler2004responsibility,triantafyllou2022actual,triantafyllou2023towards,halpern2018towards,alechina2020causality}, among which we focus on those operationalized in Dec-POMDPs~\cite{chockler2004responsibility,tigard2021artificial,triantafyllou2023towards}. However, we note that prior research has also considered alternative approaches to responsibility attribution. Baier et al.~\cite{baier2021game} propose a game-theoretic framework for responsibility attribution, distinguishing between two notions of responsibility: forward (prescriptive) and backward (retrospective). The latter is related to causal notions of responsibility~\cite{chockler2004responsibility} and attributes responsibility for a realized play, whereas the former attributes responsibility based on all potential plays. Hence, it is similar to the notion of blame in multi-agent Markov Decision Processes proposed by~\cite{triantafyllou2021blame}, which considers average performance rather than realized outcomes when attributing blame. 
Similar notions of forward- and backward-looking responsibility have been explored under other formal frameworks~\cite{yazdanpanah2019strategic}.
Moreover, some works on causal responsibility distinguish between responsibility, blame, and blameworthiness (e.g., see~\cite{chockler2004responsibility, halpern2018towards}), incorporating agents’ epistemic states and intentions into formal models. Much prior work quantifies responsibility (or related concepts) using concepts from cooperative game theory, such as the well-known Shapley value~\cite{shapley201617,shapley1954method} and Banzhaf index~\cite{banzhaf1964weighted,banzhaf1968one}. That said, alternative approaches have also been explored, such as computational models of responsibility judgments based on counterfactual simulation~\cite{tsirtsis2024towards}. Our contribution to this line of work is to understand the alignment between formal models of responsibility based on actual causality and human responsibility judgments. 


{\bf Moral responsibility.}
Work on defining morality goes back centuries with philosophers such as David Hume and Jeremy Bentham offering some of the earliest systematic accounts of moral judgments in the mid-18th century \cite{hume2023treatise,morris2001david}. Historically, much of this normative philosophical work centered on the principle of harm avoidance--the idea that moral action requires avoiding and preventing harm, whether directly or indirectly. Other moral considerations, such as honesty (e.g., refraining from lying, deception, or breaking promises), were often considered secondary when they conflicted with the imperative to prevent harm \cite{mill2001utilitarianism,bentham1988principles}. 

A fundamental aspect of morality involves making judgments about when individuals are morally responsible for their actions and holding them accountable for the resulting consequences. 
While an agent's moral responsibility may differ from their causal responsibility (for example, a toddler causing an adverse outcome may be causally responsible but not morally responsible), the two are typically closely intertwined \cite{sep-moral-responsibility}. 
This interplay between causal responsibility and moral responsibility becomes especially salient in an era of increasing human-machine interaction. 
As autonomous systems take on greater roles in decisions that may affect people's lives, understanding how society perceives responsibility for adverse outcomes 
has become increasingly vital. 
Although responsibility and blame are distinct, with philosophically \textit{interpersonal blame} considered a response to moral agents on the basis of their adverse actions \cite{sep-blame}, research on how people assign blame offers valuable insight into broader perceptions of responsibility.

\textbf{Human attitudes and perceptions.}
Prior research shows that when harm is caused by only one agent, people blame machines more than humans for the same mistake \cite{hong2020artificial,franklin2021blaming}. Similarly, robots are blamed more than humans when they fail to make utilitarian decisions \cite{malle2014theory,lima2021conflict}. However, this trend reverses in settings involving human-robot interaction when the machine operates under human supervision (such as a semi-autonomous vehicle operating under the supervision of its human driver). In such cases, people blame the human more than the machine \cite{awad2020drivers,beckers2022drivers}.
Furthermore, autonomous systems are judged more harshly than non-autonomous agents that need to rely on human input \cite{furlough2021attributing,kim2006should}. Across both cases, however, the human was assigned more blame than the agent. 

Lima et al. \cite{lima2021human} investigate people's perception of responsibility in human-AI decision-making to assess whether they align with different notions of moral responsibility. They present vignettes in which a human judge is advised either by AI tool or another human judge regarding whether to grant a defendant bail. The judge follows the given advice. Respondents are then asked to rate their agreement with various statements. The authors find that AI agents and human advisors are attributed similar levels of causal responsibility and blame for the same task. However, there was a meaningful difference in the type of responsibility: human advisors were assigned greater degrees of present- and forward-looking notions of responsibility. 

Recent work explores how people attribute blame and causal responsibility in settings where human and AI agents act independently yet contribute jointly to an adverse outcome \cite{meier2025s}. Meier et al. presented respondents with vignettes in which two agents performed identical actions that together led to an adverse outcome. No harm would have occurred if either one had refrained from acting. However, their framework defined one agent as violating an expected norm by performing that action, while the other agent was not, and manipulated whether the norm-breaking agent was a human or a machine. 
The authors found that the norm-violating agent was consistently perceived as significantly more blameworthy and causally responsible than the norm-conforming agent, and the agent type (human vs machine) did not have an effect.

Our work extends this literature by examining whether formal definitions of responsibility attribution align with human judgments in settings where humans collaborate with AI agents in multi-agent, sequential decision-making scenarios. Specifically, we study a decision-making context more complex than those explored in prior work: humans and AI agents make sequential choices, rendering the attribution of responsibility for an adverse outcome (such as the loss of a game) substantially more intricate. 
Moreover, this work is the first to examine how formal models of responsibility based on actual causality align with human judgments.



\section{Methodology}
Our survey was based on a version of the game called {\em Goofspiel} \cite{triantafyllou2022actual, lanctot2019openspiel, meng2023efficient}, which we used to simulate interactions between two competing teams, each comprising one human and one AI player. We conducted a survey experiment with respondents recruited via the Prolific platform. Respondents were asked to evaluate agent responsibility across a range of carefully designed Goofspiel scenarios or vignettes. We first describe \textit{Goofspiel} and the vignettes shown to respondents, followed by the treatments considered in the survey design. Finally, we detail our recruitment protocol. 

\begin{figure*}[ht]
    \centering
    \includegraphics[width=0.8\textwidth]{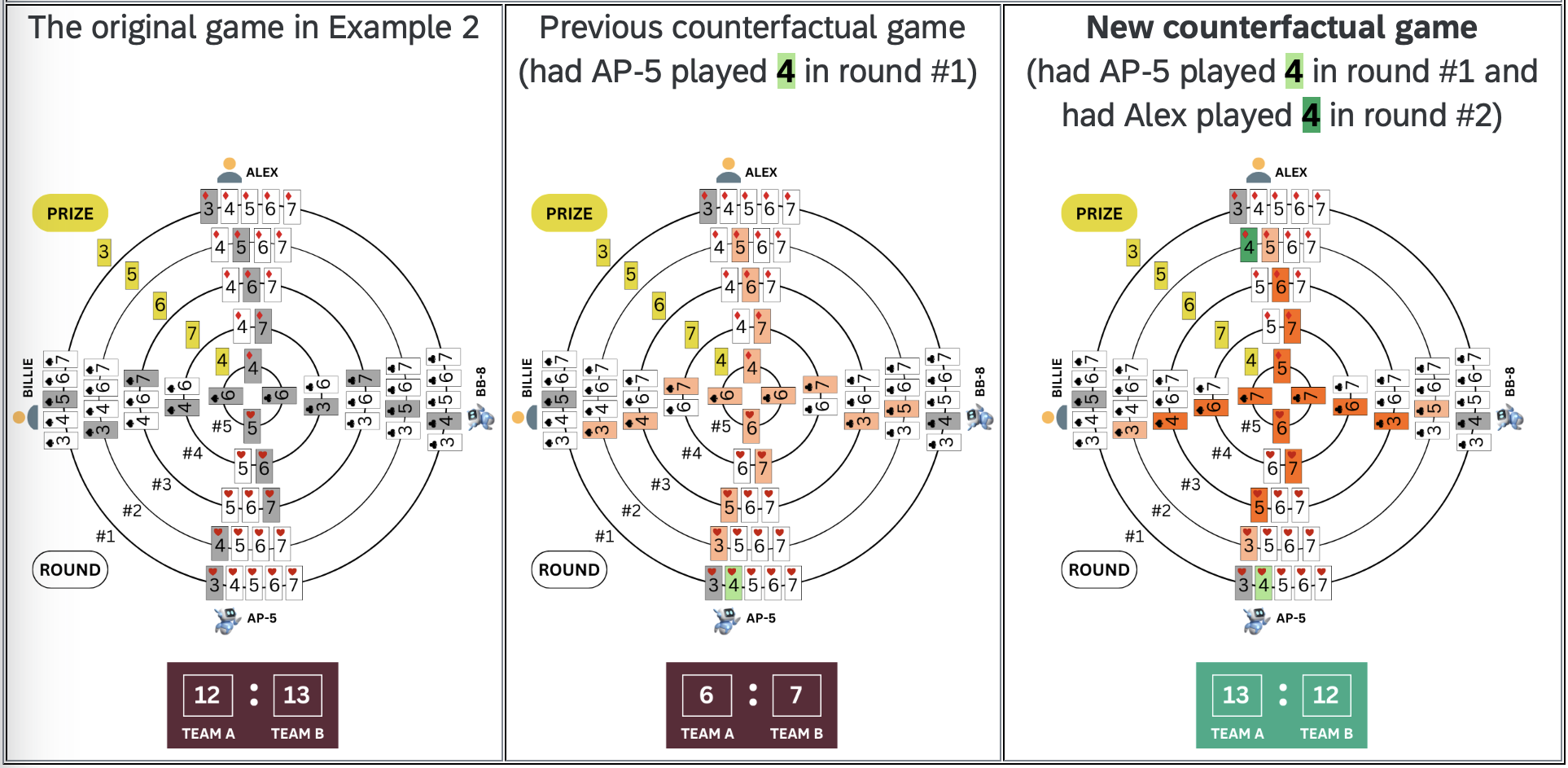} 
    \caption{An illustrative example of an original Goofspiel game and its counterfactual variants, as presented to respondents in the tutorial. In the original game (left panel), cards highlighted in gray indicate those played by the players. In the counterfactual games (middle and right panels), gray cards again mark the moves identical to those in the original game. For players holding both a gray and a light green card (agent AP-5 in round 1 of the counterfactual games), the gray card denotes the move that would have been made in the original game, while the light green card represents the intervention introduced in the counterfactual game. Cards highlighted in light orange across subsequent rounds show the moves played following this first intervention. In the case of two interventions (right panel), the light orange cards indicate the moves agents would have played after the first intervention. For players with both a light orange and a dark green card (agent Alex in round 2), the light orange card reflects the move that would have occurred without the second intervention, while the dark green card marks the second intervention itself. Cards highlighted in dark orange in later rounds represent the moves made after the second intervention.}
    \label{fig:tutorial}
\end{figure*}

\subsection{The Goofspiel game}

We consider a team-based variant of the card game \textit{Goofspiel}, following Triantafyllou et al. \cite{triantafyllou2022actual}, played over five rounds. As shown in Figure \ref{fig:tutorial}, two teams -- Team \(A\) and Team \(B\) -- compete, each composed of one human player and one AI agent.  We refer to the players in Team \(A\) as Alex (human) and AP-5 (AI). The players in Team \(B\) are referred to as Billie (human) and BB-8 (AI). The names for the human players were chosen to be gender-neutral \cite{brylla2009female}, while the names for the AI players were inspired by droid characters from the fictional Star Wars universe \cite{StarWarsAP5,StarWarsBB8}. 

Throughout the paper, we use the terms ``players'' and ``agents'' interchangeably. 
Figure \ref{fig:tutorial} depicts our game setup: the left panel (original game) shows one match, whereas the middle and right panels depict two counterfactual scenarios of that match, which we describe in more detail below. 
 
{\bf Description of Goofspiel.} At the start of the game, each player is dealt five cards from the standard 52-card deck. All players are aware that any given player holds cards of only one suit. Additionally, a separate deck of cards---called prize deck---is shuffled and placed face down. The game proceeds over five rounds. In each round, a prize card is drawn from the prize deck and revealed to all players; its value is shown in yellow between the players Alex and Billie. 
Upon observing the prize card, each player simultaneously selects one card from their hand to play. The selected card is shown in color, while the remaining unplayed cards appear in white. The team whose combined bid (i.e., the sum of the two selected cards) is higher wins the round and receives the prize card, which is added to the team's score. 
If the teams have equal sums, the prize card is discarded and no points are awarded (i.e., the scores remain the same). All cards played in that round are then discarded, and a new round begins. 
The game ends after round 5, and the team with the highest cumulative score wins. If the teams have equal scores, then the game is a draw.

{\bf Counterfactual Games.} The left panel of Figure \ref{fig:tutorial} presents the actual sequence of play. In contrast, the middle and right panels illustrate counterfactual versions of the game. In the counterfactual game in the  middle panel, AP-5 selects a  different card in Round 2. The alternative card AP-5 plays in the counterfactual is highlighted in light green, while the card originally played by AP-5 in that round originally is shown in gray. Following the literature on causality, we refer to this as an \textit{intervention} \cite{halpern2005causes,triantafyllou2022actual}.
This alters the composition of AP-5's hand, influencing subsequent rounds. In later rounds, AP-5 may sometimes choose the same card as in the original sequence and at other times a different one. All cards played after the initial intervention are highlighted in light orange, signaling both the counterfactual nature of the game and the intervention itself. At its core, a counterfactual game poses the question: \textit{What would have happened if AP-5 had played card \#4 in Round 1?}

Counterfactual games can also include multiple interventions. 
Consider the right panel of Figure \ref{fig:tutorial}. 
AP-5 plays a different card in Round 1, while Alex chooses a different card in Round 2. This scenario effectively asks: \textit{What would have happened if AP-5 had played card \#4 in Round 1 and Alex had played card \#4 in Round 2?} In such cases, the first intervention is highlighted in light green and the second in dark green. Cards played after the first intervention are shown in light orange, while those played after the second intervention are shown in dark green. As with simpler counterfactuals, in later rounds players may either replicate the card they played in the original game or select different cards. When both interventions occur in the same round, they are jointly highlighted in light green, and all subsequent plays are marked in light orange. 

Intervention(s) in counterfactual games can lead to a decisive shift in outcome. For example, in Figure \ref{fig:tutorial} (right panel), the result is reversed, with Team \(A\) winning in the second counterfactual game, unlike in the original. The comparison between the original game and its counterfactuals shows how alternative decisions by one or both agents can causally alter the game's final result. 

{\bf Game Scenario Generation.} While respondents are presented with a description of Goofspiel in which two teams—humans and AIs—play against each other, we generate Goofspiel games by rolling out policies from \cite{triantafyllou2022actual}. Specifically, for Team A, we use the policy of agent Ag$0$ from \cite{triantafyllou2022actual}, while for Team B, we use the stochastic Opponents’ policies from the same source. The initial configuration of each game depends on the {\em initial condition} treatment, which we describe in more detail in the next section. The generation process otherwise follows that of \cite{triantafyllou2022actual}: (1) before the game starts we shuffle the prize deck; (2) in each round we sample Team B's actions. With this process, we obtain $1000$ games per initial configuration in which Team A did not win for use in our survey.

Following the steps in \cite{triantafyllou2022actual}, we further identify actual causes and degrees of responsibility for each of the original games under each combination of actual causality definitions and responsibility attribution methods. In contrast to \cite{triantafyllou2022actual}, we limit the number of interventions included in counterfactual games to $2$ instead of $4$. In treatments where we show counterfactual scenarios, the set of actual causes for a given game scenario defines the counterfactual scenarios presented to respondents.
For example, in a treatment where we present one counterfactual scenario per game relevant to calculating the degree of responsibility of AP-5, we use the actual cause most relevant for determining AP-5’s degree of responsibility. In a treatment where we present two counterfactual scenarios relevant to calculating the degrees of responsibility of AP-5 and Alex, we analogously use the actual causes most relevant for determining their respective degrees of responsibility. As shown in Figure 1 (the middle panel), the tutorials for some treatments include counterfactual games that are not based on actual causes but serve as illustrative examples for explaining counterfactuals. Counterfactual games generated by actual causes change the outcome of the original game, as shown in Figure 1 (right panel).

\subsection{Survey Design: Vignettes}

The respondents first went through a tutorial explaining the rules of {\em Goofspiel}, counterfactual games, and the evaluation task they were required to perform. Note that the respondents in the survey were not provided with a description of the game scenario generation process, explained in the previous subsection. A more detailed description of the tutorial is provided in the appendix.

After the tutorial, each respondent viewed five Goofspiel game scenarios---vignettes---in which Team A loses.
Each vignette included none, one, or two different counterfactual games, depending on the aid in counterfactual reasoning treatment (described in detail in Section \ref{sec:method:treatments}). These counterfactual games, generated using actual causes as explained in the previous sections, result in Team
A winning. For each vignette shown to a respondent, the respondent was asked to rate the degree of responsibility of each Team A player (Alex and AP-5) for Team A losing the original game, using a 3-point Likert scale: \textit{Low Responsibility}, \textit{Medium Responsibility}, or \textit{High Responsibility}. They are also also asked to provide confidence in their judgments using another 3-point Likert scale: \textit{Low Confidence}, \textit{Medium  Confidence}, or \textit{High  Confidence}. Figures \ref{fig:factual_only}, \ref{fig:one_cf}, and \ref{fig:two_cf} from the appendix, show vignettes for different treatment conditions, described in the next subsection.

\subsection{Survey Design: Treatments}\label{sec:method:treatments}

To investigate factors that shape human judgments of responsibility, we focus on four key features of the decision-making context. Our experimental design varies: 
\begin{enumerate}
    \item {\bf responsibility attribution method}, i.e., the definition of actual causality and the degree of responsibility used to determine the counterfactual scenarios shown to respondents and the reference values for the degrees of responsibility of Alex and AP-5;  
    \item {\bf initial condition}, i.e., whether there is any bias in the initial hands of different players;
    \item {\bf state observability}, i.e., the information available to players during the game;
    \item {\bf aid in counterfactual reasoning}, i.e., the extent of information available to survey respondents about counterfactual alternatives.
\end{enumerate}
We use between-subjects and within-subjects manipulations across these dimensions to isolate their individual and combined effects on responsibility judgments (see Table \ref{tab:experiment_conditions} in the appendix). The following paragraphs describe our treatments: the first is within-subject, while the remaining ones are between-subject.


\textbf{Responsibility attribution method (\textsc{AC} $\times$ \textsc{DR}). } 
This treatment varies the logic used to generate counterfactual games and reference degrees of responsibility, using five distinct responsibility attribution methods, i.e., \textsc{AC} $\times$ \textsc{DR} combinations, detailed in the previous section. 
%
%
This treatment enables us to study the alignment between responsibility attribution methods based on actual causality and human judgments about responsibility. Respondents viewed one scenario per \textsc{AC} $\times$ \textsc{DR} combination, enabling a within-subjects comparison across possible \textsc{AC} $\times$ \textsc{DR} combinations. The order of the combinations was randomized for each participant. 

\textbf{Initial condition.} 
This treatment introduces asymmetries in strategic advantage by manipulating the initial distribution of card values across players, simulating real-world inequities in resource access and opportunity. In different conditions, players begin with stronger or weaker hands, affecting their ability to act effectively. 

This treatment includes four conditions. In the \textit{unbiased} condition, all players receive the same cards (of value 3-7), ensuring a level playing field. The three \textit{biased} conditions introduce targeted disadvantages. In \textit{bias against Alex}, Alex receives low-value cards (2–6), while other players receive cards (3–7). In \textit{bias against AP-5}, AP-5 is disadvantaged and receives low-value cards (2–6), while other players receive cards (3–7). In \textit{bias against both}, both Alex and AP-5 receive low-value cards (2–6), while their opponents, Team B, receive the default cards (3–7).

This treatment enables us to study how biased resource allocations shape human responsibility judgments in multi-agent sequential decision-making, particularly in mixed human–AI teams, and whether respondents distinguish between human and AI agents under unequal starting conditions. Respondents were randomly assigned to one biased condition.

\textbf{State observability.}
This treatment manipulates the perceived information structure of the game by varying what respondents are told about the information on which the players were basing their decisions. Two conditions were employed: \textit{imperfect-information} and \textit{perfect-information} games, which depict common settings in multi-agent systems \cite{shoham2008multiagent}.

In the \textit{imperfect-information} condition, respondents are told that players can see only their own hand, implying that they must decide which card to play without knowing the cards of other players.
In contrast, in the \textit{perfect-information} condition, respondents are told that all players can see every player’s hand, implying that players are more certain about the state of the game.
A note below each vignette reminds respondents of the information structure, explicitly stating whether players could see only their own cards or everyone’s cards when making their decisions.

This treatment offers a natural testbed to examine how the amount of information the players had access to during the game affects people's perceptions of responsibility in human-AI teams. It enables us to study if the uncertainty introduced due to imperfect information in the partial information setting affects responsibility attribution. Respondents were randomly assigned to one condition.

{\bf Aid in counterfactual reasoning.}
This treatment varies the amount of counterfactual information shown to respondents. We have four different conditions. 
In the \textit{factual-only} condition, respondents see only the original game. They view the players' original actions and the outcome (Team A losing). In other conditions, one or more counterfactuals are shown alongside the original:
\begin{itemize}
    \item \textit{counterfactual for Alex}: 
    respondents additionally see a counterfactual game generated from the actual cause that is most relevant for determining the degree of Alex's responsibility using the underlying \textsc{AC} $\times$ \textsc{DR} condition;
    \item \textit{counterfactual for AP-5}: respondents additionally see a counterfactual game generated from the actual cause that is most relevant for determining the degree of AP-5's responsibility using the underlying \textsc{AC} $\times$ \textsc{DR} condition;
    \item \textit{counterfactuals for both}: respondents additionally see two counterfactual games from the previous two conditions (\textit{counterfactual for Alex} and \textit{counterfactual for AP-5}).
\end{itemize}
Figure \ref{fig:factual_only} in the appendix shows the factual-only condition; Figures \ref{fig:one_cf} and \ref{fig:two_cf} add counterfactuals for one or both agents, respectively. Respondents were randomly assigned to one of these four conditions.

This treatment allows us to study how explicit counterfactuals influence perceptions of responsibility. It probes two key questions: 
\begin{enumerate}
    \item How sensitive are humans' responsibility judgments to the presence and extent of counterfactual comparisons?
    \item Do these effects differ based on whether the agent in question is human or AI?
\end{enumerate}
Varying the counterfactual(s) shown---Alex's, AP-5's, both, or none---tests if judgments shift from what \textit{did} occur to what \textit{could have occurred}, based on each agent's perceived capacity to act otherwise.

\subsection{Survey Design: Recruitment Protocol}
We conducted an a priori power analysis to determine the sample size to detect medium effects (\(\text{Cohen's d} = 0.5\)) with 80\% power at significance level \(\alpha = 0.05\). The \textsc{AC} $\times$ \textsc{DR} treatment's within-subjects design (five vignettes per participant) increased statistical power by reducing individual variability.
For all between-subjects treatments, \( \sim 20\) respondents were needed per condition, yielding a target sample of 640 across 32 unique treatment combinations.
Respondents were recruited via the Prolific, restricted to U.S.-based users with \(>95\% \) approval and \(\geq100 \) completed tasks to ensure data quality. Each was randomly assigned to one treatment combination and received US\$3.41—above U.S. federal minimum wage. This survey received approval from the ethical review board of the institution with which some of the authors are affiliated.\footnote{Due to the anonymity requirements of the AAMAS call for papers, we will provide more details in the final version of the paper.}

\section{Results}\label{sec.main}

We present our findings in two parts: how each treatment shaped respondents' \textbf{responsibility ratings} for Alex and AP-5, and the alignment of \textbf{respondents' judgments aligned} with the responsibility assignments produced by the formal \textsc{AC}x\textsc{DR} models.

\subsection{What Shapes Humans' Responsibility Judgments?}\label{sec:res:ratings}
We used a multi-variate linear mixed-effects regression model designed to predict respondents' responsibility ratings for the human agent Alex and AI agent AP-5. This allows us to assess whether the experimental treatments systematically influenced the degree of responsibility respondents assigned to each agent. The model includes fixed effects for all key experimental treatments (\textsc{AC}x\textsc{DR} combination, initial condition, state observability, aid in counterfactual reasoning), as well as respondents' self-reported confidence. 
To account for the repeated-measures nature of the design---each respondent rated five distinct Goofspiel game vignettes---we included random intercepts for both respondent ID and vignette ID, 
capturing individual- and scenario-level variability. All reported \(p-\)values reflect tests using the standard significance threshold of \(\alpha = 0.05\). Regression results are shown in Table \ref{tab:mixed_model_rating} in the appendix; below we provide an analysis of these results.

\begin{figure}[ht]
        \includegraphics[width=\columnwidth]{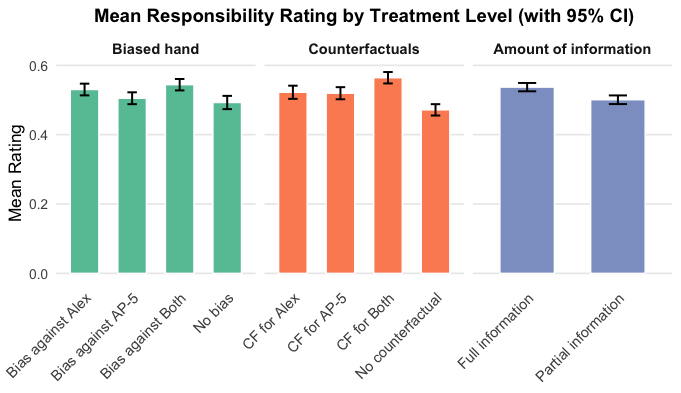}
        \caption{Mean responsibility rating with 95\% confidence intervals. Colors represent treatment types; bars within each color indicate levels of that treatment.}
        \label{fig:mean_ratings}
    \label{fig:results_summary}
\end{figure}

%
\textbf{Effect of actual causality and responsibility definitions.}
We first evaluate whether the effect that different \textsc{AC}$\times$\textsc{DR} combinations have on the respondents' responsibility rating; as noted in the previous section, \textsc{AC}$\times$\textsc{DR} determine counterfactuals shown to respondents. While no \textsc{AC}$\times$\textsc{DR} combination showed a statistically significant main effect, some (e.g., \textsc{HP}$\times$\textsc{TR}, \(p  \approx 0.09\)) exhibited marginal trends. This suggests that certain formalisms may exert a greater effect on human judgments than others, implicitly through the choice of counterfactuals shown to human raters.

\textbf{Effect of biased initial condition.} 
Bias in initial hands produced a nuanced but noteworthy effect on responsibility attributions. Responsibility ratings increased significantly when the \textit{human player, Alex, was disadvantaged} by holding worse cards \((p = 0.0182)\) and when \textit{both the human player, Alex, and the AI player, AP-5, were disadvantaged} with worse cards \((p = 0.0111)\), relative to the unbiased baseline in which all players began the game with identical cards (see also the left plot in Figure \ref{fig:mean_ratings}). By contrast, \textit{bias against AI player, AP-5, alone} did not yield a significant shift from the baseline \((p = 0.3257)\).  

These findings reveal an asymmetry in how biased resource allocation affects responsibility attribution: respondents appeared to hold human players (i.e., Alex) more accountable under conditions of disadvantage than their AI teammate. This asymmetry suggests that responsibility judgments are not only sensitive to fairness in resource distribution but also mediated by whether the disadvantaged party is human or artificial.
This pattern is consistent with prior research indicating that perceptions of fairness, agency, and obligation are differentially applied to human and AI agents \cite{awad2020drivers, lima2023blaming}.

\textbf{Effect of information available to players.}
Next, we evaluate how state observability shapes responsibility judgments. In our survey, the respondents assigned significantly greater responsibility in the full-information condition than in the partial-information condition \((p = 0.0256)\) (see also the right plot in Figure \ref{fig:mean_ratings}). 
These findings indicate that responsibility judgments are sensitive to the amount of information available to players at the time of decision-making, with access to more information leading to harsher judgments of responsibility. The results suggest that laypeople expect heightened accountability when agents are better informed and better positioned to act effectively, consistent with prior research linking transparency and access to more information to increased perceptions of accountability \cite{awad2020drivers}.

\textbf{Effect of number of counterfactuals presented.} 
Showing counterfactuals significantly increased responsibility ratings, with the strongest effect when both agents' counterfactuals were presented \((p < 0.001)\). Compared to the \textit{factual-only} baseline, responsibility ratings increased with a \textit{counterfactual for Alex} (\(p = 0.0312\)) or \textit{counterfactual for AP-5} \((p = 0.0126)\), and were highest when \textit{counterfactuals were shown for both}\((p < 0.001)\) (see also the middle plot in Figure \ref{fig:mean_ratings}). 
This demonstrates that counterfactual framing strongly shapes perceived responsibility. 
This underscores the importance of how explanations are framed in human–AI systems: what is shown or omitted can systematically bias responsibility attributions.
These findings reinforce the need for careful consideration of how hypothetical alternatives or explanations are communicated, especially when accountability is critical.

\textbf{Effect of confidence.}
We also observe a strong positive correlation between respondents' self-reported confidence and responsibility ratings \((p <  0.001)\). Respondents who expressed greater certainty in their judgments were more likely to assign higher levels of responsibility. This suggests that subjective confidence may serve as a useful proxy for perceived causal clarity---when individuals feel more confident they may also perceive the causal structure of the scenario as more determinate or interpretable.

\subsection{Alignment with Formal Responsibility Models}\label{sec:res:agreement}

We next examine the extent to which respondents' responsibility judgments align with the responsibility assignments generated by different \textsc{AC} $\times$ \textsc{DR}  combinations. For every respondent-vignette pair, we compute an \textit{agreement score} quantifying how closely their ratings of Alex and AP-5 match the responsibility levels prescribed by the  \textsc{AC} $\times$ \textsc{DR} combination used in the vignette. Agreement scores range from 0 (no alignment) to 1 (full alignment), with an intermediate score of 0.5 indicating agreement for one player but not the other. In other words, a score of 1 reflects full correspondence between respondent judgments and the \textsc{AC} $\times$ \textsc{DR} treatment-prescribed responsibilities for both players; a score of 0 reflects no such correspondence; and a score of 0.5 indicates partial agreement.

To capture how well respondents' judgments reflect the \textsc{AC} $\times$ \textsc{DR} combinations presented in the vignettes, we estimate a multivariate linear mixed-effects regression model with the agreement score as the dependent variable. Fixed effects include all experimental treatments (\textsc{AC}x\textsc{DR} combination, initial condition, state observability, aid in counterfactual reasoning), as well as respondents' self-reported confidence.
Random intercepts for respondent ID and vignette ID  
are included to account for repeated measures and scenario-specific variation. This multivariate framework allows us to isolate the impact of each treatment on alignment with formal models of responsibility attribution while appropriately modeling the nested structure of the data. 
Regression results are shown in Table \ref{tab:mixed_model_reg_new} in the appendix; below we provide an analysis.

\textbf{Effect of biased initial condition.}
We find that biased hands drive alignment. Respondents in the 
\textit{bias against both agents} 
condition showed significantly higher agreement with the AC$\times$DR treatment-prescribed responsibility attributions \((p=0.0027)\), 
while those in the \textit{unbiased} condition were significantly less aligned \((p=0.0033)\). This suggests that people not only take structural disadvantage into account when assessing responsibility, but are also more aligned with formal models of responsibility attribution when visible bias is present. In contrast, when no visible bias is present, respondents may rely more on intuitive or heuristic judgments.

\textbf{Effects of \textsc{AC}$\times$\textsc{DR} combination, available information, and number of counterfactuals presented.} 
We found no significant effects on respondents' alignment with the \textsc{AC}$\times$\textsc{DR} treatment-prescribed responsibility assignments across these three treatments--
\textsc{AC}$\times$\textsc{DR} combination, state observability, aid in counterfactual reasoning.
Limiting agents' access to information did not significantly affect agreement, nor did the selective presentation of counterfactuals meaningfully shift agreement. Finally, we found no significant differences in agreement scores across \textsc{AC}$\times$\textsc{DR} combinations--that is, no particular \textsc{AC}$\times$\textsc{DR} combination consistently aligned more closely with respondents' responsibility ratings for Alex and AP-5. This suggest that respondents did not systematically favor one responsibility attribution method over another, underscoring the difficulty of capturing human moral intuitions through formal causal frameworks. This highlights the need for further research to understand how structural, informational, and conceptual factors interact to shape human agreement with formal causal frameworks in collaborative human-AI decision-making contexts.

\textbf{Effect of confidence.} As in our prior analysis, self-reported confidence was a significant predictor of agreement \( (p=0.0015) \). Respondents who reported higher confidence were more likely to align with the ACxRD treatment-prescribed responsibility assignment. This finding further supports the idea that confidence reflects a sense of causal clarity or coherence with one's internal model of responsibility.


\section{Conclusion}\label{sec.conclusion}
To summarize, this work presents the first systematic study of how human perceptions of responsibility align with formal responsibility attribution methods grounded in actual causality. Using a large-scale survey built around a team-based version of the Goofspiel game, we examined how people assign responsibility to human and AI agents in collaborative, sequential decision-making settings. Our results reveal several key insights: a biased initial hand selectively increases perceived responsibility for the human agent, Alex, but not for the AI agent, AP-5; full-information settings lead to higher responsibility ratings; and the presence of counterfactuals substantially amplified perceived responsibility, particularly when respondents are shown counterfactuals for both agents.

We further find that alignment between human judgments and formal responsibility attribution models was strongest when structural asymmetries were present in the players' initial hands at the beginning of the game, although no 
responsibility attribution method
significantly matched human judgment across all contexts. These findings highlight the complexity of modeling accountability in sequential human–AI decision-making teams and underscore the need for more research into responsibility attribution methods that integrate both formal causal reasoning and human perceptions of responsibility and agency.


\bibliographystyle{unsrt}
\bibliography{main}

@inproceedings{tsirtsis2024towards,
  title={Towards a computational model of responsibility judgments in sequential human-AI collaboration},
  author={Tsirtsis, Stratis and Gomez Rodriguez, Manuel and Gerstenberg, Tobias},
  booktitle={Proceedings of the Annual Meeting of the Cognitive Science Society},
  volume={46},
  year={2024}
}

@book{shapley201617,
  title={17. A value for n-person games},
  author={Shapley, Lloyd S.},
  year={2016},
  publisher={Princeton University Press}
}

@article{shapley1954method,
  title={A method for evaluating the distribution of power in a committee system},
  author={Shapley, Lloyd S. and Shubik, Martin},
  journal={The American Political Science Review},
  volume={48},
  number={3},
  pages={787--792},
  year={1954},
  publisher={JSTOR}
}

@article{banzhaf1964weighted,
  title={Weighted voting doesn't work: {A} mathematical analysis},
  author={Banzhaf III, John F.},
  journal={Rutgers Law Review},
  volume={19},
  pages={317},
  year={1964},
  publisher={HeinOnline}
}

@article{banzhaf1968one,
  title={One man, 3.312 votes: a mathematical analysis of the Electoral College},
  author={Banzhaf III, John F.},
  journal={Villanova Law Review},
  volume={13},
  pages={304},
  year={1968},
  publisher={HeinOnline}
}

@article{triantafyllou2021blame,
  title={On blame attribution for accountable multi-agent sequential decision making},
  author={Triantafyllou, Stelios and Singla, Adish and Radanovic, Goran},
  journal={Advances in Neural Information Processing Systems},
  volume={34},
  year={2021}
}

@article{alechina2020causality,
  title={Causality, responsibility and blame in team plans},
  author={Alechina, Natasha and Halpern, Joseph Y. and Logan, Brian},
  journal={arXiv preprint arXiv:2005.10297},
  year={2020}
}

@inproceedings{halpern2018towards,
  title={Towards formal definitions of blameworthiness, intention, and moral responsibility},
  author={Halpern, Joseph Y. and Kleiman-Weiner, Max},
  booktitle={Proceedings of the AAAI Conference on Artificial Intelligence},
  volume={32},
  year={2018}
}

@inproceedings{yazdanpanah2019strategic,
  title={Strategic responsibility under imperfect information},
  author={Yazdanpanah, Vahid and Dastani, Mehdi and Alechina, Natasha and Logan, Brian and Jamroga, Wojciech},
  booktitle={Proceedings of the 18th International Conference on Autonomous Agents and Multiagent Systems AAMAS 2019},
  pages={592--600},
  year={2019},
}

@techreport{pearl1998definition,
  author      = {Pearl, Judea},
  title       = {On the definition of actual cause},
  type = {Technical Report R-259},
  institution = {Department of Computer Science, University of California, Los Angeles},
  year={1998}
}

@article{hitchcock2001intransitivity,
  title={The intransitivity of causation revealed in equations and graphs},
  author={Hitchcock, Christopher},
  journal={The Journal of Philosophy},
  volume={98},
  number={6},
  pages={273--299},
  year={2001},
  publisher={JSTOR}
}

@article{hall2007structural,
  title={Structural equations and causation},
  author={Hall, Ned},
  journal={Philosophical Studies},
  volume={132},
  number={1},
  pages={109--136},
  year={2007},
  publisher={Springer}
}

@article{halpern2015graded,
  title={Graded causation and defaults},
  author={Halpern, Joseph Y. and Hitchcock, Christopher},
  journal={The British Journal for the Philosophy of Science},
  volume={66},
  number={2},
  pages={413--457},
  year={2015},
  publisher={Oxford University Press}
}

@incollection{van2015moral,
  title={Moral responsibility},
  author={Van de Poel, Ibo},
  booktitle={Moral responsibility and the problem of many hands},
  pages={12--49},
  year={2015},
  publisher={Routledge}
}

@inproceedings{triantafyllou2023towards,
  title={Towards Computationally Efficient Responsibility Attribution in Decentralized Partially Observable MDPs},
  author={Triantafyllou, Stelios and Radanovic, Goran},
  booktitle={Proceedings of the 2023 International Conference on Autonomous Agents and Multiagent Systems},
  pages={131--139},
  year={2023}
}

@book{shoham2008multiagent,
  title={Multiagent systems: Algorithmic, game-theoretic, and logical foundations},
  author={Shoham, Yoav and Leyton-Brown, Kevin},
  year={2008},
  publisher={Cambridge University Press}
}

@book{hart1985causation,
  title={Causation in the Law},
  author={Hart, Herbert Lionel Adolphus and Honor{\'e}, Tony},
  year={1985},
  publisher={OUP Oxford}
}

@inproceedings{halpern2015modification,
  title={A modification of the Halpern-Pearl definition of causality},
  author={Halpern, Joseph Y},
  booktitle={Proceedings of the 24th International Conference on Artificial Intelligence},
  pages={3022--3033},
  year={2015}
}

@inproceedings{triantafyllou2022actual,
  title={Actual causality and responsibility attribution in decentralized partially observable markov decision processes},
  author={Triantafyllou, Stelios and Singla, Adish and Radanovic, Goran},
  booktitle={Proceedings of the 2022 AAAI/ACM Conference on AI, Ethics, and Society},
  pages={739--752},
  year={2022}
}

@book{halpern2016actual,
  title={Actual causality},
  author={Halpern, Joseph Y},
  year={2016},
  publisher={MiT Press}
}

@inproceedings{lima2023blaming,
  title={Blaming humans and machines: What shapes people’s reactions to algorithmic harm},
  author={Lima, Gabriel and Grgi{\'c}-Hla{\v{c}}a, Nina and Cha, Meeyoung},
  booktitle={Proceedings of the 2023 CHI conference on human factors in computing systems},
  pages={1--26},
  year={2023}
}

@article{lanctot2019openspiel,
  title={OpenSpiel: A framework for reinforcement learning in games},
  author={Lanctot, Marc and Lockhart, Edward and Lespiau, Jean-Baptiste and Zambaldi, Vinicius and Upadhyay, Satyaki and P{\'e}rolat, Julien and Srinivasan, Sriram and Timbers, Finbarr and Tuyls, Karl and Omidshafiei, Shayegan and others},
  journal={arXiv preprint arXiv:1908.09453},
  year={2019}
}

@inproceedings{meng2023efficient,
  title={An efficient deep reinforcement learning algorithm for solving imperfect information extensive-form games},
  author={Meng, Linjian and Ge, Zhenxing and Tian, Pinzhuo and An, Bo and Gao, Yang},
  booktitle={Proceedings of the AAAI Conference on Artificial Intelligence},
  volume={37},
  number={5},
  pages={5823--5831},
  year={2023}
}

@article{awad2020drivers,
  title={Drivers are blamed more than their automated cars when both make mistakes},
  author={Awad, Edmond and Levine, Sydney and Kleiman-Weiner, Max and Dsouza, Sohan and Tenenbaum, Joshua B and Shariff, Azim and Bonnefon, Jean-Fran{\c{c}}ois and Rahwan, Iyad},
  journal={Nature human behaviour},
  volume={4},
  number={2},
  pages={134--143},
  year={2020},
  publisher={Nature Publishing Group UK London}
}

@article{baier2021game,
  title={A game-theoretic account of responsibility allocation},
  author={Baier, Christel and Funke, Florian and Majumdar, Rupak},
  journal={arXiv preprint arXiv:2105.09129},
  year={2021}
}

@article{halpern2005causes,
  title={Causes and explanations: A structural-model approach. Part I: Causes},
  author={Halpern, Joseph Y and Pearl, Judea},
  journal={The British journal for the philosophy of science},
  year={2005},
  publisher={The University of Chicago Press}
}

@article{chockler2004responsibility,
  title={Responsibility and blame: A structural-model approach},
  author={Chockler, Hana and Halpern, Joseph Y},
  journal={Journal of Artificial Intelligence Research},
  volume={22},
  pages={93--115},
  year={2004}
}

@article{lima2021conflict,
  title={The conflict between people’s urge to punish AI and legal systems},
  author={Lima, Gabriel and Cha, Meeyoung and Jeon, Chihyung and Park, Kyung Sin},
  journal={Frontiers in Robotics and AI},
  volume={8},
  pages={756242},
  year={2021},
  publisher={Frontiers Media SA}
}

@article{tigard2021artificial,
  title={Artificial moral responsibility: How we can and cannot hold machines responsible},
  author={Tigard, Daniel W},
  journal={Cambridge Quarterly of Healthcare Ethics},
  volume={30},
  number={3},
  pages={435--447},
  year={2021},
  publisher={Cambridge University Press}
}

@article{brylla2009female,
  title={Female names and male names. Equality between the sexes},
  author={Brylla, Eva},
  year={2009},
  publisher={York University}
}

@misc{StarWarsBB8, 
title={List of Star Wars Characters: BB-8},
author={Star Wars Disney},
url={https://www.starwars.com/databank/bb-8}, 
journal={StarWars.com}, 
publisher={Disney}}

@misc{StarWarsAP5, 
title={List of Star Wars Characters: AP-5},
author={Star Wars Disney},
url={https://www.starwars.com/databank/ap-5}, 
journal={StarWars.com}}

@book{hume2023treatise,
  title={A treatise of human nature: Being an attempt to introduce the experimental method of reasoning into moral subjects},
  author={Hume, David},
  year={2023},
  publisher={Broadview Press}
}

@article{morris2001david,
  title={David hume},
  author={Morris, William Edward and Brown, Charlotte R},
journal={Stanford Encyclopedia of Philosophy},
  year={2001}
}

@misc{bentham1988principles,
  title={The principles of morals and legislation},
  author={Bentham, Jeremy},
  year={1988},
  publisher={New York: Prometheus Books}
}

@article{mill2001utilitarianism,
  title={Utilitarianism. Edited by George Sher},
  author={Mill, John Stuart},
  journal={Indianapolis: Hackett},
  year={2001}
}

@InCollection{sep-moral-responsibility,
	author       =	{Talbert, Matthew},
	title        =	{{Moral Responsibility}},
	booktitle    =	{The {Stanford} Encyclopedia of Philosophy},
	editor       =	{Edward N. Zalta and Uri Nodelman},
	howpublished =	{\url{https://plato.stanford.edu/archives/fall2025/entries/moral-responsibility/}},
	year         =	{2025},
	edition      =	{{F}all 2025},
	publisher    =	{Metaphysics Research Lab, Stanford University}
}

@article{malle2014theory,
  title={A theory of blame},
  author={Malle, Bertram F and Guglielmo, Steve and Monroe, Andrew E},
  journal={Psychological Inquiry},
  volume={25},
  number={2},
  pages={147--186},
  year={2014},
  publisher={Taylor \& Francis}
}

@article{furlough2021attributing,
  title={Attributing blame to robots: I. The influence of robot autonomy},
  author={Furlough, Caleb and Stokes, Thomas and Gillan, Douglas J},
  journal={Human factors},
  volume={63},
  number={4},
  pages={592--602},
  year={2021},
  publisher={Sage Publications Sage CA: Los Angeles, CA}
}

@inproceedings{kim2006should,
  title={Who should I blame? Effects of autonomy and transparency on attributions in human-robot interaction},
  author={Kim, Taemie and Hinds, Pamela},
  booktitle={ROMAN 2006-The 15th IEEE international symposium on robot and human interactive communication},
  pages={80--85},
  year={2006},
  organization={IEEE}
}

@InCollection{sep-blame,
	author       =	{Tognazzini, Neal and Coates, D. Justin},
	title        =	{{Blame}},
	booktitle    =	{The {Stanford} Encyclopedia of Philosophy},
	editor       =	{Edward N. Zalta and Uri Nodelman},
	howpublished =	{\url{https://plato.stanford.edu/archives/fall2025/entries/blame/}},
	year         =	{2025},
	edition      =	{{F}all 2025},
	publisher    =	{Metaphysics Research Lab, Stanford University}
}

@article{meier2025s,
  title={Who's to Blame? Blame Attribution in Autonomous Human-Machine Interactions},
  author={Meier, Jeremy and Wylie, Richard and Laham, Simon M},
  journal={Blame Attribution in Autonomous Human-Machine Interactions (June 16, 2025)},
  year={2025}
}

@article{franklin2021blaming,
  title={Blaming automated vehicles in difficult situations},
  author={Franklin, Matija and Awad, Edmond and Lagnado, David},
  journal={Iscience},
  volume={24},
  number={4},
  year={2021},
  publisher={Elsevier}
}

@article{hong2020artificial,
  title={Why is artificial intelligence blamed more? Analysis of faulting artificial intelligence for self-driving car accidents in experimental settings},
  author={Hong, Joo-Wha and Wang, Yunwen and Lanz, Paulina},
  journal={International Journal of Human--Computer Interaction},
  volume={36},
  number={18},
  pages={1768--1774},
  year={2020},
  publisher={Taylor \& Francis}
}

@article{beckers2022drivers,
  title={Drivers of partially automated vehicles are blamed for crashes that they cannot reasonably avoid},
  author={Beckers, Niek and Siebert, Luciano Cavalcante and Bruijnes, Merijn and Jonker, Catholijn and Abbink, David},
  journal={Scientific reports},
  volume={12},
  number={1},
  pages={16193},
  year={2022},
  publisher={Nature Publishing Group UK London}
}

@inproceedings{lima2021human,
  title={Human perceptions on moral responsibility of AI: A case study in AI-assisted bail decision-making},
  author={Lima, Gabriel and Grgi{\'c}-Hla{\v{c}}a, Nina and Cha, Meeyoung},
  booktitle={Proceedings of the 2021 CHI conference on human factors in computing systems},
  pages={1--17},
  year={2021}
}



\clearpage
\onecolumn
\appendix 
{\allowdisplaybreaks

\section{Summary of Experiment Treatments}
\begin{table}[ht]
\centering
\caption{Survey experiment treatments. `Treatment' denotes the experimental condition, while 'Randomization Variation' details whether the treatment was randomized within- or between-subjects. `Treatment Levels' denotes possible values the treatment can have, and `Signal' showcases the key features of the decision-making context the treatment enables us to study.
}
\begin{tabularx}{\textwidth}{>{\raggedright\arraybackslash}p{2.8cm} >{\raggedright\arraybackslash}p{3cm} >{\raggedright\arraybackslash}p{4.2cm} >{\raggedright\arraybackslash}X}
\toprule
\textbf{Treatment} & \textbf{Randomization Variation} & \textbf{Treatment Levels} & \textbf{Signal} \\
\midrule

\textsc{AC}x\textsc{DR} & Within-subjects &
BF × CH \newline
HP × CH \newline
HP × TR \newline
TR × CH \newline
TR × TR  &
Different \textsc{AC}x\textsc{DR} combinations \\
\addlinespace

Number of Counterfactuals (CFs) & Between-subjects &
Factual only \newline
Factual + CF for Agent 1 \newline
Factual + CF for Agent 2 \newline
Factual + CF1 + CF2 &
Counterfactuals shown to the survey respondents\\
\addlinespace

Amount of Information & Between-subjects &
Partial: Players cannot see each other’s cards \newline
Full: Players can see each other’s cards &
Visibility of other agents’ hands \\
\addlinespace

Biased Hand & Between-subjects &
Opponents: 3–7, Agent 1: 1–5, Agent 2: 5–9 \newline
All agents: 3–7 \newline
Opponents: 3–7, Agent 1: 5–9, Agent 2: 1–5 \newline
Bias against both agents &
Relative strength of hands across teams \\
\bottomrule
\end{tabularx}
\label{tab:experiment_conditions}
\end{table}

\newpage
\section{Regression Table: Respondents' Responsibility Ratings}
\begin{table}[H]
\centering
\caption{Multivariate linear mixed-effects model predicting survey respondents' responsibility judgment \textit{rating} for Alex and AP-5. Fixed effects include all treatments in the experiment (ACxDR combination, initial condition, state
observability, aid in counterfactual reasoning) and respondents' self-reported confidence. Random effects include the respondent ID and the specific game scenario shown.}
\label{tab:mixed_model_rating}
\begin{tabular}{lcccccc}
\toprule
\textbf{Fixed Effects} & \textbf{Estimate} & \textbf{Std. Error} & \textbf{df} & \textbf{t value} & \textbf{Pr(>|t|)} & \textbf{Signif.} \\
\midrule
(Intercept) & 0.3505 & 0.02631 & 1211 & 13.322 & $<$2e-16 & *** \\
agent\_factorAP5\_rating & -0.002225 & 0.007647 & 6168 & -0.291 & 0.7710 & \\
info\_factorfull & 0.03344 & 0.01495 & 740.4 & 2.237 & 0.0256 & * \\
bias\_factor0 & 0.05220 & 0.02206 & 709.6 & 2.366 & 0.0182 & * \\
bias\_factor1 & 0.02178 & 0.02214 & 711.3 & 0.984 & 0.3257 & \\
bias\_factorboth & 0.05611 & 0.02203 & 735.2 & 2.547 & 0.0111 & * \\
CF\_agent\_factor1-CF-0 & 0.04638 & 0.02148 & 750.5 & 2.159 & 0.0312 & * \\
CF\_agent\_factor1-CF-1 & 0.05218 & 0.02088 & 744.8 & 2.500 & 0.0126 & * \\
CF\_agent\_factor2-CF-both & 0.09614 & 0.02099 & 739.0 & 4.580 & 5.46e-06 & *** \\
acra\_assigned\_factorHPxCH & 0.02022 & 0.01235 & 1204 & 1.638 & 0.1017 & \\
acra\_assigned\_factorHPxTR & 0.02101 & 0.01236 & 1095 & 1.699 & 0.0895 & . \\
acra\_assigned\_factorTRxCH & 0.02069 & 0.01235 & 1236 & 1.675 & 0.0941 & . \\
acra\_assigned\_factorTRxTR & 0.01144 & 0.01235 & 1168 & 0.926 & 0.3544 & \\
confidence & 0.1490 & 0.01605 & 4668 & 9.288 & $<$2e-16 & *** \\
\midrule
\textbf{Random Effects} & \textbf{Group} & \textbf{Name} & \textbf{Variance} & \textbf{Std. Dev.} & & \\
\midrule
 & PROLIFIC\_PID\_factor & (Intercept) & 0.02974 & 0.17245 & & \\
 & trajectory\_factor & (Intercept) & 0.00051 & 0.02249 & & \\
 & Residual &  & 0.10510 & 0.32420 & & \\
\midrule
\multicolumn{7}{l}{REML criterion at convergence: 5276.9} \\
\multicolumn{7}{l}{Number of observations: 7190; Groups: PROLIFIC\_PID\_factor (714), trajectory\_factor (498)} \\
\bottomrule
\multicolumn{7}{l}{\textit{Scaled residuals:} Min = -2.53935;\quad 1Q = -0.69025;\quad Median = 0.01037;\quad 3Q = 0.68597;\quad Max = 2.48765} \\
\multicolumn{7}{l}{\textit{Signif. codes:} *** $p<0.001$, ** $p<0.01$, * $p<0.05$, . $p<0.1$}
\end{tabular}
\end{table}

\newpage
\section{Regression Table: Alignment with Formal Responsibility Models}

\begin{table}[H]
\centering
\caption{Multivariate linear mixed-effects model predicting \textit{agreement\_value}, or the agreement score quantifying how closely respondents' ratings for Alex and AP-5 matched the responsibility levels prescribed by the underlying ACxDR combination. Fixed effects include all treatments in the experiment (ACxDR combination, initial condition, state
observability, aid in counterfactual reasoning) and respondents' self-reported confidence. Random effects include the respondent ID and the specific game scenario shown.}
\label{tab:mixed_model_reg_new}
\begin{tabular}{lcccccc}
\toprule
\textbf{Fixed Effects} & \textbf{Estimate} & \textbf{Std. Error} & \textbf{df} & \textbf{t value} & \textbf{Pr(>|t|)} & \textbf{Signif.} \\
\midrule
(Intercept) & 0.3024 & 0.0290 & 1162 & 10.427 & $<$2e-16 & *** \\
info\_factorpartial & -0.0122 & 0.0159 & 705.7 & -0.769 & 0.4424 & \\
bias\_factor1 & -0.0071 & 0.0218 & 685.1 & -0.324 & 0.7462 & \\
bias\_factorboth & 0.06698 & 0.0223 & 692.4 & 3.007 & 0.0027 & ** \\
bias\_factornone & -0.06796 & 0.0231 & 667.1 & -2.947 & 0.0033 & ** \\
CF\_agent\_factor1-CF-1 & 0.02843 & 0.0223 & 705.3 & 1.273 & 0.2033 & \\
CF\_agent\_factor2-CF-both & 0.02364 & 0.0224 & 673.3 & 1.057 & 0.2908 & \\
CF\_agent\_factorfactual-none & 0.00875 & 0.0230 & 720.0 & 0.380 & 0.7038 & \\
acra\_assigned\_factorHPxCH & -0.00576 & 0.0198 & 1282 & -0.291 & 0.7714 & \\
acra\_assigned\_factorHPxTR & -0.00898 & 0.0199 & 1171 & -0.451 & 0.6519 & \\
acra\_assigned\_factorTRxCH & 0.01194 & 0.0200 & 1286 & 0.598 & 0.5503 & \\
acra\_assigned\_factorTRxTR & -0.00972 & 0.0200 & 1218 & -0.486 & 0.6269 & \\
confidence & 0.06684 & 0.0210 & 2158 & 3.189 & 0.00145 & ** \\
\midrule
\textbf{Random Effects} & \textbf{Group} & \textbf{Name} & \textbf{Variance} & \textbf{Std. Dev.} & & \\
\midrule
 & PROLIFIC\_PID\_factor & (Intercept) & 0.01645 & 0.12824 & & \\
 & trajectory\_factor & (Intercept) & 0.00476 & 0.06898 & & \\
 & Residual &  & 0.11702 & 0.34208 & & \\
\midrule
\multicolumn{7}{l}{Number of observations: 3422; Groups: PROLIFIC\_PID\_factor (714), trajectory\_factor (484)} \\
\multicolumn{7}{l}{REML criterion at convergence: 2913.7} \\
\bottomrule
\multicolumn{7}{l}{\textit{Signif. codes:} *** $p<0.001$, ** $p<0.01$, * $p<0.05$}
\end{tabular}
\end{table}

\newpage
\section{Tutorials Shown to Survey Respondents}
Depending on the counterfactual treatment assigned, respondents viewed either a tutorial presenting only the factual scenario or one incorporating counterfactuals. Figure \ref{fig:tut-fact-1} displays the first screen of the tutorial for the factual-only condition. After introducing the basic setup of the Goofspiel game, respondents proceeded to the next screen (Figure \ref{fig:tut-fact-2}), which explained the game rules in detail. On this screen, participants were required to answer several comprehension questions about the cards played, prizes, and round outcomes to ensure they understood the rules. They could not advance until all responses were correct. Finally, respondents completed a brief practice vignette to reinforce the tutorial content.

\begin{figure}[H]
  \centering
  \includegraphics[width=0.8\linewidth]{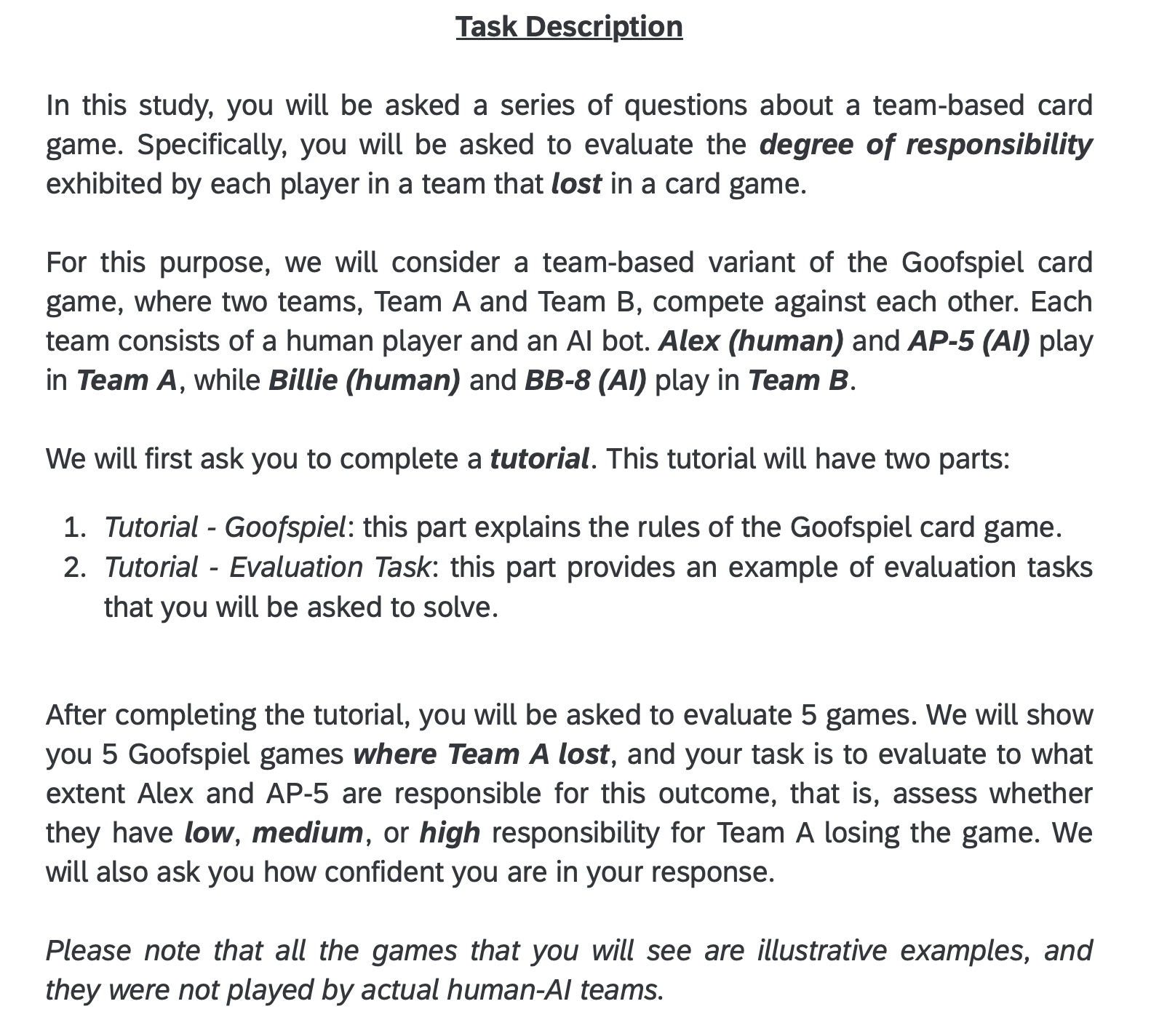}
  \caption{The first screen of the tutorial shown to participants assigned to a factual-only treatment condition.}
  \label{fig:tut-fact-1}
\end{figure}

\begin{figure}[H]
  \centering
  \includegraphics[scale=0.9]{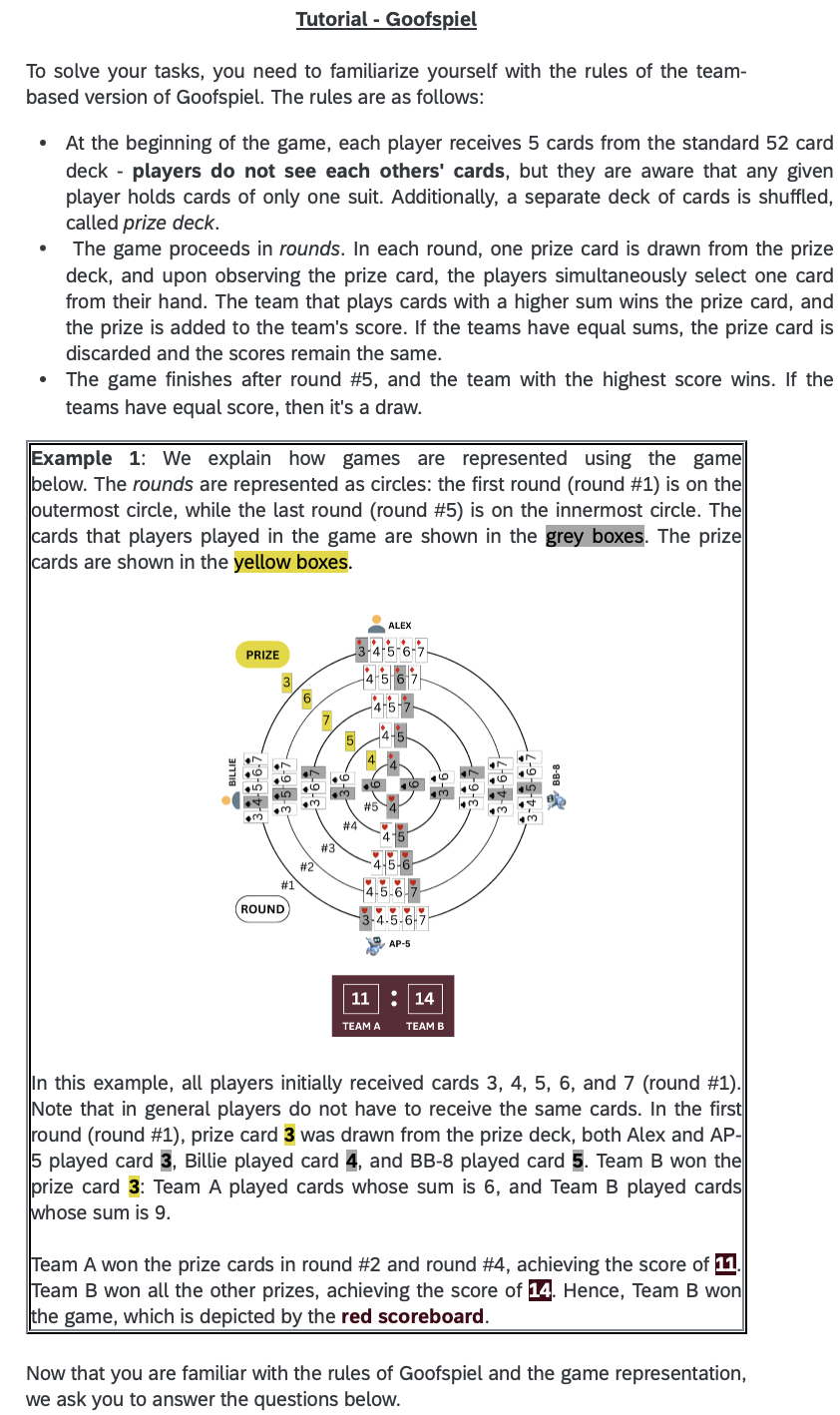}
  \caption{The second screen of the tutorial shown to participants assigned to a factual-only treatment condition. After the rules of \textit{Goofspiel} are explained, the respondents are asked to answer some questions to confirm their understanding of the game.}
  \label{fig:tut-fact-2}
\end{figure}


\begin{figure}[H]
  \centering
  \includegraphics[width=0.8\linewidth]{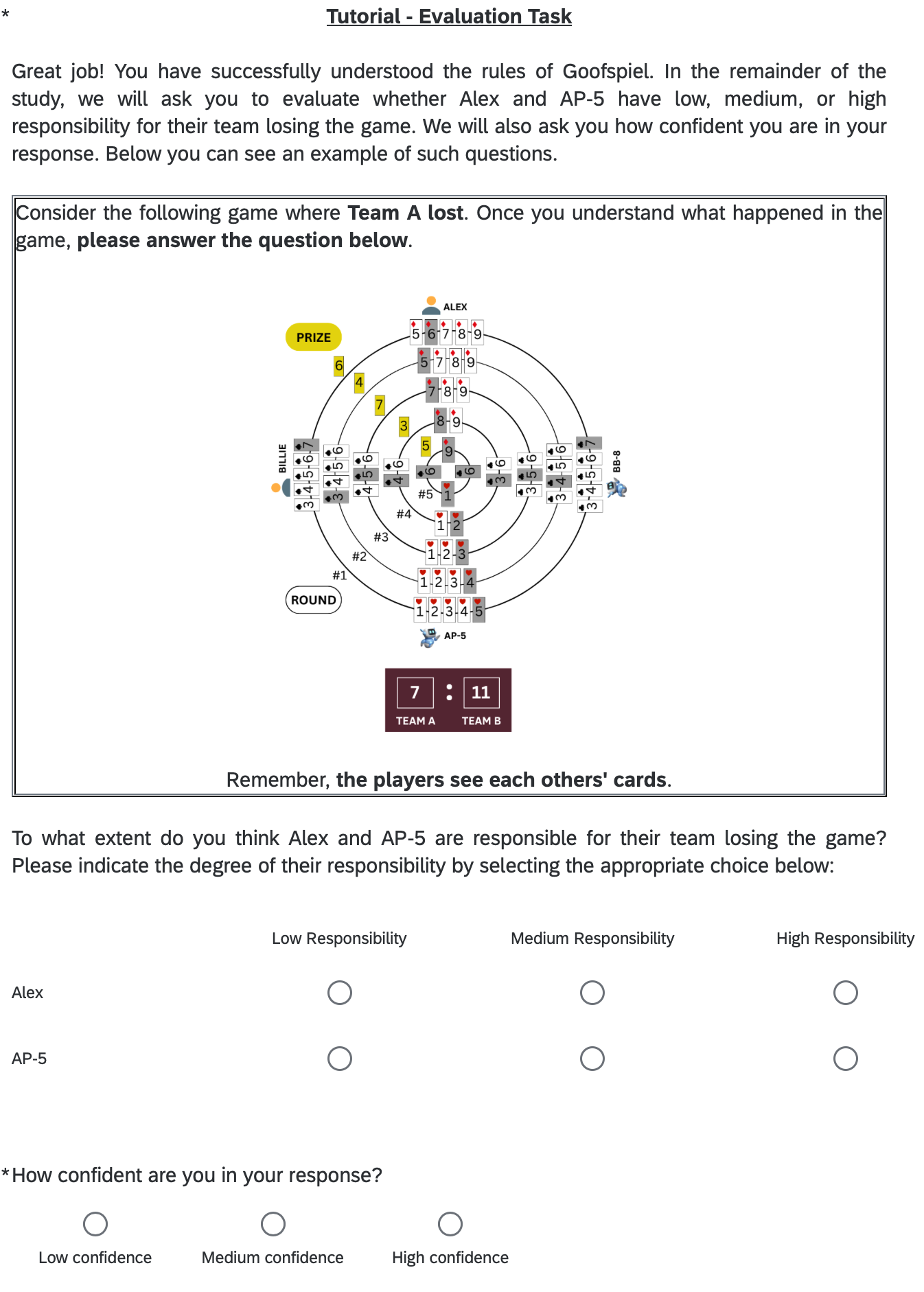}
  \caption{At the end of the tutorial, the respondent is asked to answer a practice vignette.}
  \label{fig:tut-fact-3}
\end{figure}

\section{Sample Questions Shown to Respondents}

\begin{figure}[H]
    \centering
    \includegraphics[width=0.8\textwidth]{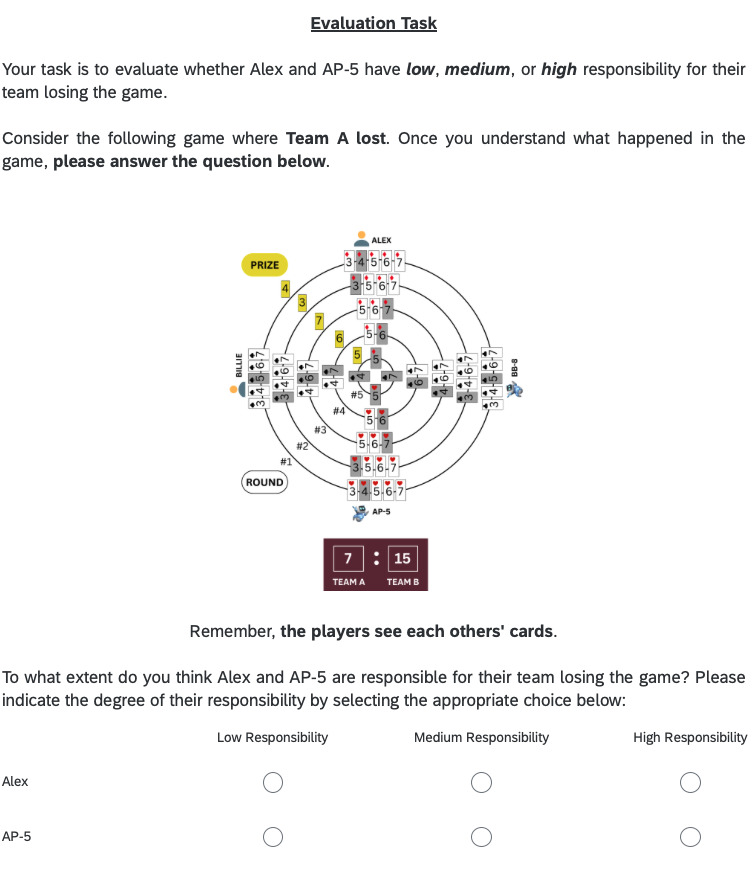} 
    \caption{An illustrative example of a Goofspiel game shown to survey respondents under full information settings with no counterfactuals.}
    \label{fig:factual_only}
\end{figure}


\begin{figure}[H]
    \centering
    \includegraphics[width=0.8\textwidth]{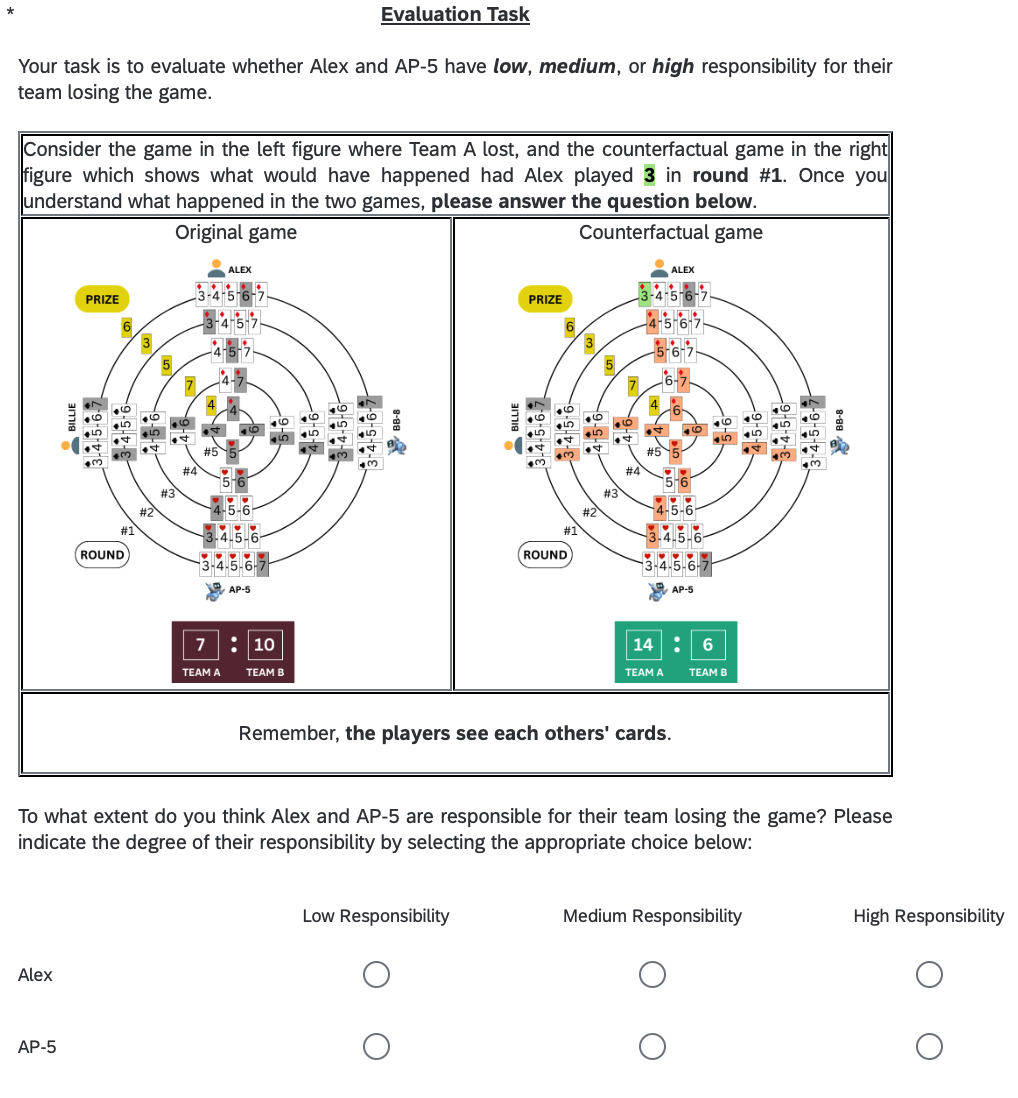} 
    \caption{An illustrative example of a Goofspiel game shown to survey respondents under full information settings and the counterfactual for one agent.}
    \label{fig:one_cf}
\end{figure}


\begin{figure}[H]
    \centering
    \includegraphics[width=0.8\textwidth]{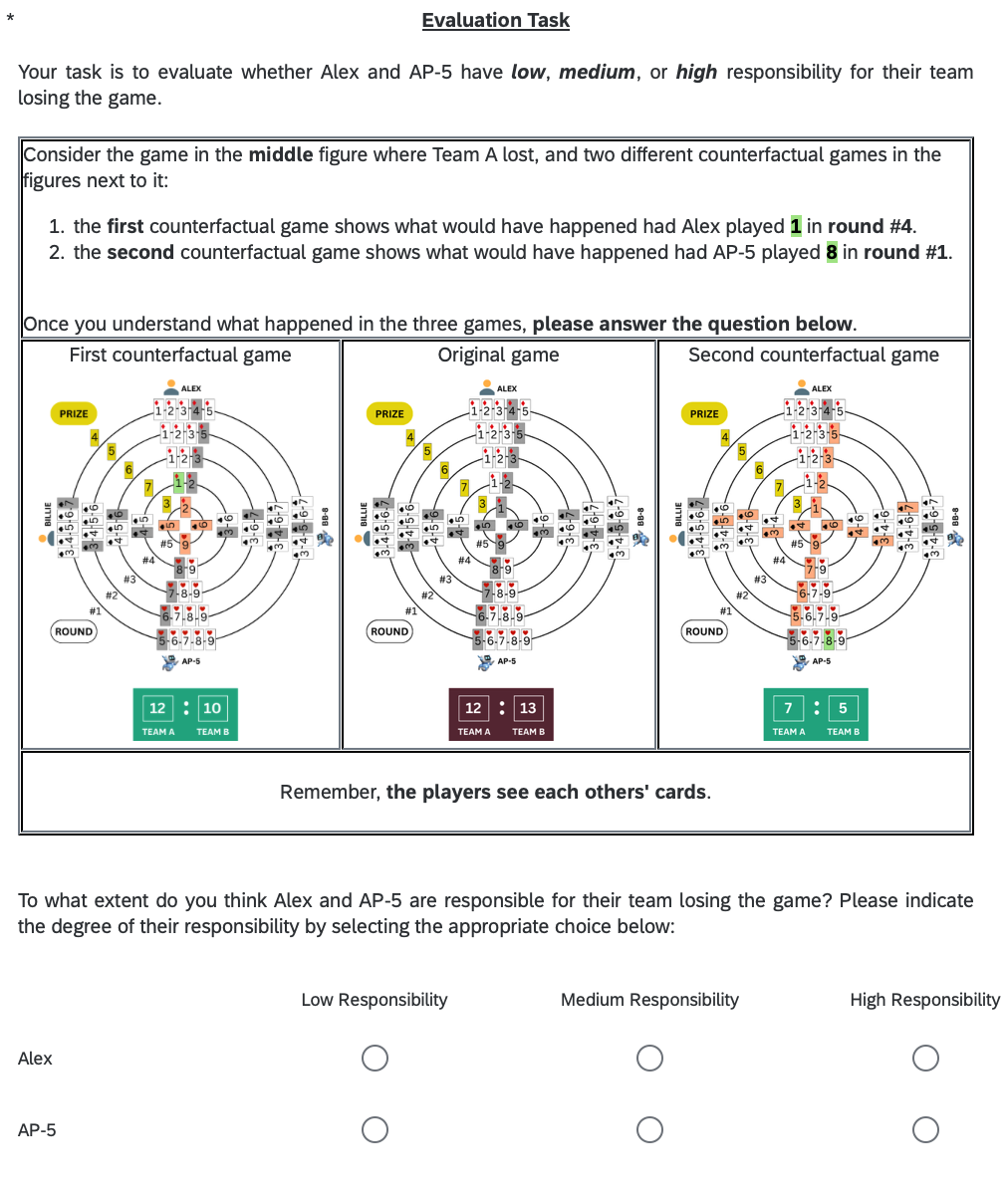} 
    \caption{An illustrative example of a Goofspiel game shown to survey respondents under full information settings and the counterfactuals for both agents.}
    \label{fig:two_cf}
\end{figure}

}
{}
\end{document}